\documentclass[aps,preprint,nofootinbib]{revtex4-1}
\usepackage{graphicx}
\usepackage{array}
\begin{document}
\title{\Large \bf Black Strings in  Ho\v{r}ava-Lifshitz Gravity}
\author{\large Alikram N. Aliev and \c{C}etin  \c{S}ent\"{u}rk}
\address{Feza G\"ursey Institute, \c Cengelk\" oy, 34684   Istanbul, Turkey}
\date{\today}

\begin{abstract}

We examine a  class of cylindrically  symmetric solutions  in  Ho\v{r}ava-Lifshitz gravity. For the relativistic value of the coupling constant, $\lambda=1 $, we find  the ``hedgehog" type static black string solution with the  nonvanishing radial shift in the ADM-type decomposition of  the spacetime metric. With zero radial shift, this solution corresponds to the usual BTZ  black string in general relativity. However, unlike  the general relativity case,  the BTZ  type black strings do naturally  exist in HL gravity, without the need for any specific source term.  We also find  a rotating BTZ type black  string solution which requires the  nonvanishing radial shift for its very existence. We calculate the mass and the angular momentum of this solution, using the canonical Hamiltonian approach. Next, we discuss the Lemos type black string, which is inherent in general relativity  with a negative cosmological constant, and present the static metric for any value of $ \lambda > 1/3 $. Finally,  we show that while, for $\lambda=1 $, the entropy of the Lemos type black string is given by  {\it one quarter} of the  horizon area, the entropy of the static BTZ type black string  is {\it one half} of its  horizon area.

\end{abstract}


\maketitle

\section{introduction}

Recently, Ho\v{r}ava  put forward the idea of gravity endowed with Lifshitz-type anisotropic scaling \cite{hora1, hora2}. This is an intriguing attempt to formulate a consistent quantum field theory of gravity in $ 3+1 $  dimensions by invoking the  anisotropy  between space and time, first introduced in condensed matter systems  \cite{lif}. The degree of the anisotropy given  by a number  $ z \,$,  the ``dynamical critical exponent", plays the role of an important observable in the theory, determining its behavior at short scales.  The  Ho\v{r}ava-Lifshitz (HL) theory of gravity exhibits an anisotropic scaling with $ z=3 $  fixed point at short distances, thereby becomes  a power-counting renormalizable  in the ultraviolet (UV) regime. Thus, in this approach the classical theory of gravity acquires UV completion, being driven to a quantum field theory of nonrelativistic gravitons in $ 3+1 $  dimensions. Meanwhile, at long distances the scaling becomes isotropic, flowing to $ z=1 $, and the theory restores its relativistic invariance in the infrared (IR) regime where it resembles, through some relevant deformations, many familiar features of general relativity.

Due to its fundamentally  nonrelativistic nature, HL gravity admits a natural description in terms of the ADM-type variables, appearing in the $ 3+1 $ foliation of the spacetime metric in general relativity. These variables form triplet which consists of the spatial metric as a dynamical field, the lapse function and the shift vector. However, unlike in general relativity, the privileged  role of time in HL gravity leads to a ``preferred foliation" of spacetime by slices of constant time. Consequently, the full spacetime  symmetries of the theory  reduce to  time reparametrization  symmetry (space-independent)  and  spatial diffeomorphisms (time-dependent), which preserve the spacetime foliation. Clearly, the lapse function and the shift vector can be viewed as two gauge fields of the foliation-preserving diffeomorphisms. This fact is also encoded in the physical spectrum of the theory around flat spacetime where an extra scalar polarization of the graviton  appears. With the foliation-preserving diffeomorphisms one can naturally assume that the lapse is a function of time alone, while  the shift is a spacetime field, thereby fitting  the ``projectable"  theory of foliation \cite{hora2}.
Altogether, these properties  form a minimal basis for the realization of anisotropic scaling in gravity.  The minimal realization  also involves the concept of  the ``detailed balance" condition. This implies that the potential term in the action is effectively a square of a pre-potential, appearing in a one dimension fewer Euclidean theory. In further developments, to improve the physical content of the theory,  both the projectability  condition and the detailed balance condition were relaxed in a number of cases (see a review \cite{muko}, for details). Moreover, it was shown that an  extension of the foliation-preserving diffeomorphisms by  an Abelian gauge symmetry, eliminates the scalar polarization of the graviton that  appeared in the minimal realization of the idea of anisotropic scaling  \cite{hora3}.

Among possible applications of HL gravity, its  phenomenological  consequences  in our universe are of great importance. It is interesting that the theory results in a new  mechanism for scale-invariant cosmological perturbations, even without inflation \cite{kk, muko1}. The early history of the universe is also significantly changed with HL gravity which admits  regular cyclic and bouncing solutions \cite{kk, calc, wang1}. However, it should be emphasized that HL gravity  suffers from a number of inconsistency problems  as well. For instance,  the scalar mode becomes unstable in the UV regime \cite{calc} when keeping the detailed balance condition, but abandoning the projectability condition. There also exist scalar instabilities in the IR regime \cite{bog}, which may result  in  strong coupling problems \cite{ sibir1, wu, pang}. Furthermore, scale-invariant perturbations \cite{gao} are generated provided that the detailed balance condition is broken in the UV regime \cite{roy}. Another issue is the existence of black hole solutions. In \cite{lmp}, it was shown that the theory admits a static and spherically symmetric AdS type black hole solution. The asymptotic behavior of this solution is essentially different from that of the Schwarzschild-AdS black hole  in  general relativity. Meanwhile, the counterpart of the usual asymptotically flat  Schwarzschild  solution was found in  \cite{ks} by a relevant deformation of the HL action. This solution turned out to be very useful to figure out the observational consequences of HL gravity in both weak  and strong gravity regimes \cite{roman, harko}. Further, these type of solutions,  as well as their certain extension in the framework of the most general  spherically symmetric ansatz, were studied in \cite{cco1, cco2, lkm, wang2, dario, nijo, eyal}. As for the rotating counterparts of these solutions, they still remain unknown. In a recent work \cite{as}, some progress in this direction was achieved in the limit of slow rotation (see also Ref.\cite{lee}).

In this paper, we  examine  a  class of cylindrically  symmetric solutions in HL gravity, which can be thought of as counterparts of black strings in general relativity. In Sec.II we begin by describing the physical  content of HL gravity using the  ADM-type decomposition of  the spacetime metric and present the equations of motion underlying the theory. In Sec.III  we discuss
the general stationary and  cylindrically symmetric ansatz for spacetime metric. Focusing on the spacetimes,  for which the Cotton tensor in the HL action vanishes, we delineate two intriguing examples of the cylindrically  symmetric spacetimes which are   the counterparts of those for the BTZ  and Lemos types  black strings in general relativity \cite{lemos1, lemos2}.  The BTZ  black  strings in general relativity  are obtained by adding an extra spacelike flat dimension to the metric of the three-dimensional BTZ black hole \cite{btz}. Next, for $\lambda=1 $,  we discuss the static BTZ type black string  solutions with zero and nonzero radial shift.  In the latter case, we call it the {\it hedgehog} type solution. In this section, we also present the stationary and cylindrically  symmetric solution that describes the BTZ  type rotating black string in HL gravity. This solution is of  a hedgehog type as well, since the radial ``hair"  is inevitable to support the rotational dynamics.  We  calculate the mass and the angular momentum of this solution, employing the canonical Hamiltonian approach. We further discuss the Lemos type black string and  present the corresponding static solution for any value of the coupling constant $ \lambda > 1/3 $.  In  Sec.IV we examine the thermodynamical properties of the static black string configurations in HL gravity using the Euclidean path integral approach.

\section{Basics of  Ho\v{r}ava-Lifshitz Gravity}

The privileged role of time in HL gravity with Lifshitz type anisotropic scaling  makes it fundamentally nonrelativistic  and results in a preferred foliation of spacetime by slices of constant time. As a consequences of this, the full spacetime symmetries of the system reduce to the foliation-preserving diffeomorphisms which are generated by
\begin{equation}
\label{diff}
t\rightarrow \tilde{t}(t)\,,~~~~x^i\rightarrow\tilde{x}^i(t,x^i)\,.
\end{equation}
With this in mind, it natural to employ the ADM-type $ 3+1 $ decomposition of the spacetime metric. We have
\begin{eqnarray}
ds^2 & = &-N^2 dt^2+ g_{ij}\left(dx^i+ N^i dt\right)\left(dx^j + N^j
dt\right),
\label{admform}
\end{eqnarray}
where  the three-dimensional spatial metric  $ g_{ij} $ is a dynamical field, the lapse function $ N $ and the shift vector $ N^i $ play the role of gauge fields of  diffeomorphisms  (\ref{diff}) and therefore one can suppose that they respect the same  functional dependence. That is, the lapse is only the function of time, N= N(t),  while the shift is a spacetime function, $ N^i= N^i(t, x^i) $.  We recall that  such a  decomposition  of the spacetime metric corresponds to the ``projectable"   version of the HL gravity.

With the metric decomposition in (\ref{admform}),  the usual Einstein-Hilbert action decomposes as
\begin{equation}
I_{EH}=\frac{1}{16\pi
G}\int dt d^3x \sqrt{g}\, N \left(K_{ij}K^{ij}-K^2+R-2\Lambda\right),
\label{ehact}
\end{equation}
where $ G $ is the gravitational constant, $ K_{ij} $ is the extrinsic curvature,  $ R= g^{ij}R_{ij} $ is the  Ricci scalar,  $ \Lambda $ is the cosmological constant and
\begin{equation}
K_{ij}=\frac{1}{2N}\left(\dot{g}_{ij}-D_i N_j-D_j N_i\right),~~~~~K=g^{ij}K_{ij}\,,~~~~~N_i = g_{ij} N^j\,.
\label{ext}
\end{equation}
Here the dot denotes the derivative with respect to time  and $ D $ is
the derivative operator  with respect to the spatial metric $ g_{ij}\, $.

The action  governing the dynamics of  HL  gravity with the detailed balance condition is given by (see Ref.\cite{hora2})
\begin{eqnarray}
I &= &\int dt d^3x \sqrt{g}\, N \left\{g_0\left(K_{ij}K^{ij}-\lambda K^2 \right) + g_1 \left(R - 3 \Lambda_W\right) + g_2 R^2 + g_3 Z_{ij} Z^{ij}\right\},
\label{actHL}
\end{eqnarray}
where, for further convenience,  we have used the notations $ Z_{ij}= C_{ij}+g_4 R_{ij} $,
\begin{eqnarray}
g_0&=&\frac{2}{\kappa^2}\,,~~~g_1=\frac{\kappa^2\mu^2\Lambda_W}{8(1-3\lambda)}\,,~~~
g_2=\frac{\kappa^2\mu^2(1-4\lambda)}{32(1-3\lambda)}\,,~~~
g_3=-\frac{\kappa^2}{2\omega^4}\,,~~~g_4=- \frac{\mu \omega^2}{2}\,.
\label{g0123}
\end{eqnarray}
We note that $ \kappa $, $ \lambda $, $ \mu $ and $ \omega $ are coupling constants of the theory,  $ \Lambda_W $ is a three-dimensional cosmological constant. The Cotton tensor $ C^{ij} $ is  symmetric, traceless and covariantly constant and it is given by
\begin{equation}
C^{ij}=\frac{\epsilon^{ikl}}{\sqrt{g}}\,D_k \left(R^j_{~l}-\frac{1}{4} \delta^j_{~l}R \right),
\label{cot}
\end{equation}
where $ \epsilon^{ikl} $ is the usual Levi-Civita symbol. From action (\ref{actHL}) it follows that in HL gravity the speed of light, the Newtonian constant and the cosmological constant appear as emergent quantities. Indeed, taking the IR limit of this action, where the quadratic in curvature terms are omitted, and rescaling  the time coordinate as $ t  \rightarrow c t $,  we compare  the result with the Einstein-Hilbert action in (\ref{ehact}). This yields the emergent relations
\begin{equation}
c=\frac{\kappa^2\mu}{4}\sqrt{\frac{\Lambda_W}{1-3\lambda}}\,\,,
~~~~~~G=\frac{\kappa^2c^2}{32\pi}\,,~~~~~~ \Lambda=\frac{3}{2}\,\Lambda_W\,.
\label{emergent}
\end{equation}
In what follows, we shall focus only on the case of a negative cosmological constant. Then, from the emergent relation for the speed of light, it follows that the dynamical coupling constant of HL gravity $ \lambda $ must obey the inequality $ \lambda > 1/3 $. We shall also take $ c=1 $, without lose of generality.

The equations of motion that follow from  action  (\ref{actHL}) were obtained in \cite{lmp, kk}. Variation of the action with respect to the lapse $ N $ yields the Hamiltonian constraint
\begin{equation}
\label{ham}
-g_0 \left(K_{ij}K^{ij}-\lambda K^2\right)+ g_1 \left(R- 3\Lambda_W\right)
+g_2\, R^2 + g_3\,Z_{ij}Z^{ij}=0\,,
\end{equation}
and its variation with respect to the shift  $ N^i $  gives us  the momentum constraint
\begin{equation}
\label{mom}
D_j \left(K^{ij}-\lambda g^{ij}K\right)=0\,.
\end{equation}
Meanwhile, variation of the action with respect to the dynamical variable  $g^{ij}$ yields the equation of motion given by
\begin{equation}
\label{dyneq}
E_{ij}\equiv  g_0 \left(E^{(1)}_{ij}-\lambda  E^{(2)}_{ij}\right)
+g_1 E^{(3)}_{ij}
+g_2E^{(4)}_{ij}+g_3\left( g_4 E^{(5)}_{ij}
+E^{(6)}_{ij}\right)=0\,,
\end{equation}
where
\begin{eqnarray}
E^{(1)}_{ij}&=& 2 N_{(i}D_{|k|}K^k_{~j)}-2K^k_{~(i}D_{j)}N_k- N^k D_k K_{ij}
\nonumber \\[2mm]  & &
-2 N K_{ik} K^k_{~j}-\frac{1}{2}\,g_{ij} N K_{kl}K^{kl}+N K K_{ij}+\dot{K}_{ij}\,,\nonumber \\[2mm]
E^{(2)}_{ij}&=&\left(\frac{1}{2}NK^2-N^k\partial_k K+\dot{K}\right) g_{ij}+2 N_{(i}\partial_{j)}K\,, \nonumber \\[2mm]
E^{(3)}_{ij}&=&\left[R_{ij}-\frac{1}{2}g_{ij}\left(R-3\Lambda_W\right)
-D_i D_j+g_{ij} D^2
 \right]N,\nonumber \\[2mm]
E^{(4)}_{ij}&=& 2\left(R_{ij}-\frac{1}{4}g_{ij}R-D_i D_j+g_{ij} D^2 \right) N R \,,\nonumber \\[2mm]
E^{(5)}_{ij}&=&-2 D_k D_{(i}[Z_{j)}^{~k}N]+D^2(N Z_{ij})+g_{ij}D_k D_l(N Z^{kl})\,,\nonumber \\[2mm]
E^{(6)}_{ij}&=&\left(-\frac{1}{2}g_{ij}Z_{kl}Z^{kl}+2Z_{ik}Z^k_{~j}-2Z_{k(i}C_{j)}^{~k}+g_{ij}Z_{kl}C^{kl}\right)N
\nonumber \\[2mm] &&
-D_k[T^{kl}_{~~(i}R_{j)l}] +R^n_{~l}D_n[T^{kl}_{~~(i}g_{j)k}] -D^n[T^{kl}_{~~n}g_{k(i}R_{j)l}]\nonumber \\[2mm] &&
-D^2 D_k[T^{kl}_{~~(i}g_{j)l}] +D^n[g_{l(i}D_{j)}]D_kT^{kl}_{~~n}
+D_l D_{(i}D_{|k|}T^{kl}_{~~j)}\nonumber \\[2mm] &&
+g_{ij}D^nD_k D_lT^{kl}_{~~n}\,.
\label{dyndecom}
\end{eqnarray}
We note that in these expressions $ D^2 = D_i D^i $, \, $T^{ij}_{~~k}=N (\sqrt{g})^{-1}\, \epsilon^{ijl}Z_{lk} $ and  round parentheses over indices  denote a symmetrization procedure. Despite the fact that these  equations look  rather complicated, the authors of work \cite{lmp, kk} were the first  to find the simple  exact solutions to them, using  a standard spherically symmetric ansatz for the spacetime metric. In further developments, these type of solutions were also studied in the framework of the most general  spherically symmetric metric ansatz (see, for instance, Refs \cite{ wang2, dario}).

\section{Black String Solutions}

In this section, we discuss a class of exact cylindrically  symmetric solutions  to HL gravity. We begin with the general stationary and  cylindrically symmetric metric ansatz in the form
\begin{eqnarray}
ds^2&=&\left(-\tilde{N}^2f+N_rN^r+N_\phi N^\phi\right)dt^2+2\left(
N_r dr + N_\phi d\phi \right) dt \nonumber\\
&&+ f^{-1} dr^2+r^2d\phi^2+ g dz^2\,,
\label{metansatz}
\end{eqnarray}
where all the metric functions are assumed to depend on the radial coordinate $ r $ alone and we have redefined the lapse function as  $ N= \tilde{N}  \sqrt{f} $ for further convenience. The shift vector $ N_i= g_{ij} N^j = \{ N_r\,, N_\phi\,, 0\} $  and the three-dimensional spatial metric possesses  cylindrical symmetry, involving the functions $ f=  f(r) $ and $ g= g(r) $. We note that, just like in the spherically  symmetric case \cite{dario}, the presence of the radial shift in  metric (\ref{metansatz}) is inherent in HL gravity as the foliation-preserving invariance of the theory is not enough for eliminating it from the metric. That is, in contrast to  general relativity, in HL gravity cylindrically symmetric metrics with $ N_r=0 $ and $ N_r\neq 0 $ are not physically equivalent.

In order to simplify the consideration, we  focus on the solutions for which the Cotton tensor (\ref{cot}) vanishes. It is straightforward to show that with  metric ansatz (\ref{metansatz}), the only nonvanishing component of this tensor is given by
\begin{eqnarray}
C_{\phi z}&=&\frac{\sqrt{f}}{8rg^{5/2}}\left\{r g^2 f''\left(r g'-2g\right)+g f'\left(2g^2-2r^2g'^{\,2} + 3 r^2 g g''\right) \right. \nonumber \\[2mm]  & & \left.
+2 f \left[r^2g'^{\,3}-r^2 (g^2)'g''-g^2\left(g'-r g''-r^2g'''\right)\right]\right\}\,.
\label{cotcom1}
\end{eqnarray}
Here and in what follows, the prime denotes differentiation with respect to $ r $.  For $ g= const$,  this expression takes the most simple form
\begin{equation}
C_{\phi z}=\frac{\sqrt{f g }}{4r}(f'-r f'')
\end{equation}
and the equation $ C_{\phi z}=0 $ is immediately  solved by
\begin{equation}
f=\eta r^2- m\,,
\label{cotcomsim}
\end{equation}
where $ \eta $ and $ m $ are constants of integration. Thus, for  constant  $ g $ and  with $ f $ given in (\ref{cotcomsim}) the Cotton tensor  $ C_{ij} =0 $. For these solutions, as seen from metric (\ref{metansatz}), one can set  $ g=1 $  by  rescaling  of the $z$-coordinate.  Furthermore, we see that the resulting metric, with $  N_r=0  $,  matches  the form of the stationary BTZ black  string spacetime in general relativity \cite{lemos1} (see also Ref.\cite{ah} for the BTZ string in Cherns-Simon  gravity). We recall that  the BTZ  black  string configurations are obtained by adding an extra flat dimension to the metric of the three-dimensional BTZ black hole \cite{btz} and they require for their very existence a specific source term in the corresponding field equations. Using this analogy, we will call the class of solutions with $ g=1 $ and  $ N_r\neq 0  $, the ``BTZ type black string"  solutions. Below, we will see that that the BTZ type black string  solutions  do naturally  exist in HL gravity (without the need for any specific source term).

Meanwhile, it is not difficult to see that for $ g= \alpha^2 r^2 $, where  $ \alpha $ is a constant parameter, expression (\ref{cotcom1}) vanishes identically, irrespective of the form of the function $ f(r) $.  That is, we again have  $ C_{ij} =0 $.  In this case,  the metric ansatz in (\ref{metansatz}), with $  N_r=0  $, matches the form which describes  the stationary black string  solution of general relativity,  found by Lemos  \cite{lemos2}. This type of string solutions are inherent in general relativity with a negative cosmological constant. Below, we will also discuss the Lemos type static string solution in HL gravity  for any value of the coupling constant $\lambda > 1/3 \,$.

\subsection{BTZ type solutions}

We now need to substitute metric (\ref{metansatz}), with $ g=1 $,
into the field equations of HL gravity. In doing this, we find that the Hamiltonian constraint (\ref{ham}) takes the form
\begin{eqnarray}
&&2\eta^2 r^6 N_r^2 f^{-1} (\lambda-1)
+2 r^2 f \left[(\lambda-1)(N_r^2+r^2 N_r'{^2})+2\lambda r N_r N_r'\right]\nonumber\\[2mm] &&
+ \,\frac{\kappa^4\mu^2r^4\tilde{N}^2}{8(1-3\lambda)}\left[(2\lambda-1)\eta^2
-2\eta\Lambda_W-3\Lambda_W^2\right]-(2 N_\phi-r N_\phi')^2
\nonumber\\[2mm] &&
+ 4 \eta r^4 N_r[\lambda N_r+(\lambda-1)r N_r']= 0\,,
\label{ham1}
\end{eqnarray}
and the momentum constraint (\ref{mom}) reduces to the following two equations
\begin{eqnarray}
\label{mom1}
&&\eta r^2 f^{-2}(\lambda-1)
\left\{\eta r^2\tilde{N} N_r+ f\left[r N_r \tilde{N}'-2\tilde{N}(N_r+r N_r')\right]\right\}\nonumber\\[2mm]
&& + r\tilde{N}'[\lambda N_r+(\lambda-1)\,r N_r']
+(\lambda-1)\tilde{N}\left[N_r-r(N_r'+r N_r'')\right]=0\,,
\\[3mm]
&&\tilde{N}'(2N_\phi-rN_\phi')
-\tilde{N}(N_\phi'-rN_\phi'')= 0\,.
\label{mom2}
\end{eqnarray}
Meanwhile, calculations show that the nontrivial components of equation  (\ref{dyneq}) are given by
\begin{eqnarray}
&&\frac{\kappa^4\mu^2r^3\tilde{N}^2}{1-3\lambda}\left\{\left[(2\lambda-1)\eta^2
-2\eta\Lambda_W-3\Lambda_W^2\right]r\tilde{N}+2 f \left[(2\lambda-1)\eta-\Lambda_W\right]\tilde{N}'\right\}\nonumber \\[2mm]&&
+ 16 \eta r^5  f^{-1} N_r^2 (\lambda-1)\,\left(3\eta r\tilde{N}+2 f \tilde{N}'\right) +16 r^2 f \left\{2 r N_r \tilde{N}'\left[\lambda N_r+(\lambda-1)r N_r'\right]
\right. \nonumber \\[2mm]  & & \left.
+ (\lambda-1)\left(3N_r^2+r^2 N_r'^{\,2}\right)\tilde{N}
+ 2 r \tilde{N}  N_r\left[N_r'-(\lambda-1)r N_r''\right]\right\}
\nonumber \\[2mm] &&
-8\tilde{N}\left\{(2N_\phi- r N_\phi')^2+4\eta r^4 N_r \left[(\lambda-2)N_r+(\lambda-1)r N_r' \right]\right\}=0\,,~~~~~~ (E_{rr}=0)\,,
\label{deq1}
\end{eqnarray}
\begin{eqnarray}
&&-\frac{\kappa^4\mu^2r^4\tilde{N}^2}{1-3\lambda}\left\{\left[(2\lambda-1)\eta^2
-2\eta\Lambda_W-3\Lambda_W^2\right]\tilde{N}
+2\left[(2\lambda-1)\eta-\Lambda_W \right] \left(3\eta r\tilde{N}'+ f \tilde{N}''\right)\right\} \nonumber \\[2mm] &&
+ 16 \eta^2 r^6  f^{-1}\tilde{N} N_r^2 (\lambda-1)
-32\, r \tilde{N}'\left[N_\phi(2N_\phi-rN_\phi')-\lambda \eta r^4 N_r^2\right]
\nonumber \\[2mm] &&
-8\tilde{N}\left\{3\left(2N_\phi-rN_\phi'\right)^2+4\eta r^4 N_r\left[(3\lambda-2)N_r+(1+3\lambda)r N_r'\right]
-4 r N_\phi \left(N_\phi'-rN_\phi''\right)\right\} \nonumber \\
&&+16r^2 f \left\{2r N_r\tilde{N}'\left[(\lambda-1)N_r+\lambda r N_r'\right] +\tilde{N}\left[(\lambda-1)N_r^2-(\lambda+1)r^2 N_r'^2\right]
\right. \nonumber \\[2mm]  & & \left.
-2r N_r \tilde{N}\left[2(\lambda-1)N_r'
+\lambda r N_r''\right]\right\}=0\,, ~~~~~~  (E_{\phi\phi}=0)\,,
\label{deq2}
\end{eqnarray}
\begin{eqnarray}
&&\frac{\kappa^4 \mu^2 r^3 \tilde{N}^2 f}{1-3\lambda}\left\{ \left[(1+2\lambda)\eta^2+6\eta\Lambda_W+3\Lambda_W^2\right]r\tilde{N} +2\left(\lambda\eta+\Lambda_W\right)\
\left[(3\eta r^2 +f)\tilde{N}'+ r f \tilde{N}''\right]\right\}
\nonumber \\[2mm] &&
+ 32\lambda  r^3  f  N_r \tilde{N}' \left[(\eta r^2+f) N_r
+ r f  N_r'\right]
- 8\tilde{N}\left\{4 f N_\phi(N_\phi-rN_\phi')+2\left[(1-\lambda) m^2
\right. \right. \nonumber \\[2mm]  & & \left. \left.
+2\eta r^2 f(1+2\lambda) \right]r^2 N_r^2
+r^2 f \left[N_\phi'^{\,2}+2(\lambda+1)r^2f  N_r'^{\,2}\right]
 \right. \nonumber \\[2mm]  & & \left.
+4 r^3 f  N_r \left[(\eta r^2 +5\lambda\eta r^2 -2 \lambda m)N_r'+\lambda r f N_r''\right]\right\}= 0\,, ~~~~~~  (E_{zz}=0)\,,
\label{deq3}
\end{eqnarray}
\begin{eqnarray}
&&
N_\phi \left\{ \eta r^2 f^{-2}(\lambda-1)\left(\eta r^2\tilde{N}N_r+ f \left[rN_r\tilde{N}'-2\tilde{N}(N_r+rN_r')\right]\right)
 \right. \nonumber \\[2mm]  & & \left.
+ r\tilde{N}'\left[\lambda N_r+(\lambda-1)rN_r'\right]
+(\lambda-1)\tilde{N}\left[N_r-r(N_r'+rN_r'')\right]\right\}=0 \,,~~~~~~  (E_{r\phi}=0)\,,
\label{deq4}
\end{eqnarray}
\begin{eqnarray}
&&
\left(8\eta^2r^4-4\eta mr^2- m^2\right)\tilde{N}'
+r f \left[(7\eta r^2- m)\tilde{N}''+r f\tilde{N}'''\right]= 0 \,,~~~~~~  (E_{z \phi}=0)\,.
\label{deq5}
\end{eqnarray}
We recall that the function $ f $, appearing in these equations is given in
(\ref{cotcomsim}). For the general value of $\lambda $  these equations look somewhat complicated, but they  are drastically simplified for the relativistic value  $\lambda=1 $. To make further consideration more illustrative, it is fitting to begin with the special cases and then go up to the general case.

(i) {\it  The static  solution} ($ N_r=0 $ and $  N_\phi= 0 $). In this case,  from the Hamiltonian  constraint (\ref{ham1}), we find that
\begin{equation}
\eta =\frac{1\pm\sqrt{6\lambda-2}}{2\lambda-1}\,\Lambda_W\,.
\label{eta}
\end{equation}
whereas, the momentum constraint equations (\ref{mom1}) and (\ref{mom2}) are trivially satisfied. Taking this value of $ \eta $ into account in  equation (\ref{deq1}), we immediately fix the function $ \tilde{N}$ as  $ \tilde{N}=\tilde{N}_0 $, where  $\tilde{N}_0 $ is a constant of integration. With these quantities in mind, it is easy to check the remaining equations of motion. As a consequence, we  find that equation  (\ref{deq3}) takes the simple form given by
\begin{equation}
\frac{(1+2\lambda)(3\lambda-1)\pm(4\lambda-1)\sqrt{6\lambda-2}}{(1-2\lambda)^2}=0\,.
\label{deq3s}
\end{equation}
All other equations are satisfied automatically.  Solving equation (\ref{deq3s})  for  $ \lambda > 1/3 $, which is the case in our consideration, we  arrive at the relativistic value  $ \lambda =1 $. This value corresponds to the lower sign in  equations (\ref{eta}) and (\ref{deq3s}). In other words, starting with the general value of $ \lambda > 1/3 $,  we are driven, by the equations of motion, to the value $ \lambda =1 $.

Thus, the static and cylindrically  symmetric solution for $ \lambda =1 $  is given by
\begin{equation}
ds^2=- f dt^2 +f^{-1}dr^2 + r^2 d\phi^2+ dz^2\,,
\label{btzstat}
\end{equation}
where the metric function $ f= -\Lambda_W r^2- m $. Furthermore, we have set $\tilde{N}_0=1 $,  by  adjusting the time coordinate,  and used equations (\ref{cotcomsim}) and (\ref{eta}). It is easy to see that this metric  can be interpreted as describing the spacetime of a static BTZ string in HL gravity. It possesses  an event horizon located at the radius
\begin{equation}
r_+ =\sqrt{\frac{ m}{-\Lambda_W }}\,,
\label{btzhor1}
\end{equation}
where the quantity $ m $ plays the role of a  mass   parameter and  $ m > 0 $.

As we have mentioned  above, the BTZ black string configurations do not exist in general relativity  without introducing a specific source term into  the Einstein field equations. It is remarkable that a particular higher derivative structure of HL gravity provides a natural place for the BTZ type  black strings in this theory.  We note that  this solution  was also discussed in \cite{ck}.

(ii) {\it  The static hedgehog solution}. This is the  general static and cylindrically symmetric spacetime  with the nonvanishing radial  shift, $ N_r\neq 0 $. We  present now this solution  for $ \lambda =1 $.  From the momentum constraint equation (\ref{mom1}) we see that  the quantity $ \tilde{N} $  again remains constant, i.e. $ \tilde{N}=\tilde{N}_0 $.  Since  $ N_\phi=0 $  as well,  equations (\ref{mom2}), (\ref{deq4}) and (\ref{deq5})
become trivial. Meanwhile,   the Hamiltonian constraint (\ref{ham1}) gives
\begin{equation}
N_r=\pm f^{-1/2}\,\sqrt{\xi+\frac{\kappa^4 \mu^2}{64} \tilde{N}_0^2 r^2
\left(\eta-3\Lambda_W\right)\left(\eta+\Lambda_W\right)}\,\,,
\label{Nr1}
\end{equation}
where $ \xi $ is a constant of integration. It is not difficult to  verify that  solution (\ref{Nr1}) is also  subject to equations  (\ref{deq1}) and (\ref{deq2}). On the other hand, substituting this solution in equation (\ref{deq3}), focusing on the case when the associated spacetime metric for $ \xi =0 $ goes over into that given in (\ref{btzstat}), we find that
\begin{equation}
 \eta=-\Lambda_W\, .
\label{eta1}
\end{equation}
Finally, we arrive at the spacetime metric in the form
\begin{equation}
ds^2=- \left(\tilde{N}_0^2 - N_r^2\right)f dt^2+2 N_r dr dt+f^{-1}dr^2 + r^2 d\phi^2+ dz^2\,,
\label{btzNr}
\end{equation}
where the metric functions are given by
\begin{equation}
f= -\Lambda_W r^2- m\,, ~~~~~ N_r= \pm\sqrt{\xi/ f}\,.
\label{f2}
\end{equation}
For  $ \xi \neq 0 $, the solution describes  the BTZ type static black string  with the radial hair, i.e. the black string with  a hedgehog  behavior. Taking $ \tilde{N}_0= 1 $, we find that the horizon radius,  at which $ g^{rr}= 0 $,
is given  by
\begin{equation}
r_+ =\sqrt{\frac{m + \xi}{-\Lambda_W }}\,\,.
\label{btzhor1}
\end{equation}
We see that the radial hair contributes to the mass parameter. Clearly, the quantity $ m +\xi $   must be positive.

(iii) {\it The stationary hedgehog solutions}.  Using equations of motion given in (\ref{ham1})-(\ref{deq5}), it is straightforward to show that with the vanishing radial shift,  $ N_r=0 $,  HL gravity does not support the stationary and cylindrically symmetric solution, the rotating BTZ type black string.
However, such a solution does exist in the general stationary and  cylindrically symmetric case (with $ N_r\neq 0 $), whereby inevitably behaving as a hedgehog type solution. Turning now to this solution for $ \lambda= 1 $, we first note that the quantity  $ N_\phi $ does not enter in equation (\ref{mom1}) at all. Therefore, as in the static case, this equation  gives us  $ \tilde{N}=\tilde{N}_0 $.  With this in mind, from equation (\ref{mom2}) we find that
\begin{equation}
N_\phi=\sigma r^2+\gamma\,,
\label{Nphi}
\end{equation}
where $ \sigma $ and $ \gamma $ are constants of integration.  Substituting now this expression  in equation (\ref{ham1}) and solving it, we find that
 \begin{equation}
N_r=\pm f^{-1/2}\,\sqrt{\xi+\frac{\kappa^4 \mu^2}{64} \tilde{N}_0^2 r^2
\left(\eta-3\Lambda_W\right)\left(\eta+\Lambda_W\right)- \frac{\gamma^2}{r^2}}\,\,.
\label{Nr3}
\end{equation}
Straightforward calculations show  that with  $ \tilde{N}=\tilde{N}_0 $,  the expressions  in (\ref{Nphi}) and (\ref{Nr3}) solve all the remaining field equations, provided that the relation  in (\ref{eta1}) holds. Altogether, these expressions enable us to write down the spacetime metric in the form
\begin{eqnarray}
ds^2 &= & - \left(\tilde{N}_0^2  - N_r^2 - r^{-2} f^{-1} N_\phi^2 \right) f  dt^2 +2\left(N_r dr+2N_\phi d\phi \right) dt \nonumber\\[2mm]
&&+ f^{-1} dr^2 + r^2d\phi^2+ dz^2\,,
\label{stheg}
\end{eqnarray}
where
\begin{equation}
N_r= \pm f^{-1/2} \sqrt{\xi- \frac{\gamma^2}{r^2}}\,,
 \label{Nrf}
\end{equation}
and the functions $ f $ and $ N_\phi $ are given in equations (\ref{f2}) and  (\ref{Nphi}), respectively. Here the constant parameter $ \gamma $
can be thought of as a rotation parameter and, as  seen from equation (\ref{Nrf}), it necessarily requires  the nonvanishing radial hair,  $ \xi \neq 0 $.

The horizon structure of this solution is determined  by the equation $ g^{rr}= 0 $ and we have
\begin{eqnarray}
-\Lambda_W r^2- m- \xi +\gamma^2/r^2&=& 0\,.
\label{horeq}
\end{eqnarray}
The two two roots  of this equation, $ r_ {+} $ and  $ r_ {-} $
are given by
\begin{equation}
r_\pm^2=\frac{m+\xi}{- 2\Lambda_W}\left[1\pm\sqrt{1-\frac{4\gamma^2 (-\Lambda_W)}{(m+\xi)^2}}\,\right],
\label{stbtzhor1}
\end{equation}
and provided that
\begin{equation}
m+\xi > 0\,,~~~~~~ |\gamma| \leq \frac{m+\xi}{2\sqrt{-\Lambda_W}}\,,
\label{horconds}
\end{equation}
they give the radii of outer and inner horizons, respectively. In the extreme limit of rotation, where the equality in the second expression in (\ref{horconds}) holds, the outer and inner horizons coincide and we find that
\begin{equation}
r_ {+}^2 = r_ {-}^2 =
\frac{m+\xi}{- 2\Lambda_W}\,.
\end{equation}
The Hawking temperature  can be calculated using the standard formulae
\begin{eqnarray}
T &=&\frac{\kappa}{2\pi}= \sqrt{-\frac{1}{2}(\nabla_\mu\chi_\nu)(\nabla^\mu\chi^\nu)}\,\,,
\label{ht1}
\end{eqnarray}
where $ {\kappa} $ is the surface gravity  and the Killing  vector $\, \chi=\partial_t+\Omega_H\partial_\phi \,$ describes the isometry of the horizon that rotates with the angular velocity $ \Omega_H=- \gamma/r_+^2\, $. Performing explicit calculations, we find that
\begin{eqnarray}
T&=& -\frac{r_+^2- r_-^2}{2\pi r_+}\, \Lambda_W\,.
\label{ht2}
\end{eqnarray}
We see that in the limit of extreme rotation the Hawking temperature vanishes, just as for the extreme BTZ black hole in three-dimensional general relativity \cite{btz}.

Next, we calculate the physical mass and angular momentum  of solution  (\ref{stheg}), using  the canonical Hamiltonian  formalism \cite{btz} in HL gravity  \cite{cco1, cco2, as}.   It is straightforward to show that in this approach the action  in (\ref{actHL}) takes the form
\begin{equation}
\label{actHAM}
I=\int dt d^3x(\pi^{ij}\dot{g}_{ij}-N\mathcal{H}-N^i\mathcal{H}_i)+B\,,
\end{equation}
where
\begin{eqnarray}
\pi^{ij}&=& g_0 \,\sqrt{g}(K^{ij}-\lambda Kg^{ij})\,,\nonumber \\[2mm]
\mathcal{H}&=&\sqrt{g}\left\{g_0 (K_{ij}K^{ij}-\lambda K^2)- g_1 \left(R-3 \Lambda_W\right) - g_2 R^2 - g_3 Z_{ij}Z^{ij}\right\},\nonumber \\[2mm]
\mathcal{H}_i &=& -2 D_j\pi_i^{~j}\,,
\label{can1}
\end{eqnarray}
and $ B $  denotes a boundary term. Evaluating this action  for the metric in (\ref{stheg}) and taking  the result  per unit length of the string, we find that
\begin{equation}
\label{actden}
\mathcal{I}=- 2\pi (t_2-t_1) \int dr\left(\tilde{N} \sqrt{f}\,\mathcal{H}+N^r \mathcal{H}_r+N^\phi \mathcal{H}_\phi \right) + \mathcal{B},
\end{equation}
where
\begin{eqnarray}
\label{hquant1}
&&\mathcal{H}=\frac{1}{\sqrt{f}}\left\{g_1 \left(3\Lambda_W r+f'\right)-\frac{2 g_2+g_3 g_4^2}{2 r}\,f'^{\,2}-\frac{g_3}{8 r^3} f (f'-r f'')^2
\right. \nonumber \\[2mm]  & & \left.
+\,\frac{g_0  r^3}{2}\left(\frac{{N^\phi}{\,'}}{\tilde{N}}\right)^2
-\frac{g_0}{\tilde{N}^2}\left(\frac{{N^r}^{\,2}}{f}\right)'\right\},\\[2mm]
&&\mathcal{H}_r= 2g_0 f^{-1} \,\frac{N^r \tilde{N}'}{\tilde{N}^2}\,,\\[2mm]
&&\mathcal{H}_\phi = g_0\left(\frac{r^3{N^\phi}'}{\tilde{N}}\right)'.
\end{eqnarray}
We note that,  with the $z$-coordinate being noncompact, physically meaningful quantities are those taken per unit length of the string.  Varying this action with respect to the associated fields and omitting the terms which vanish when the equation of motion hold, we arrive at the expression
 \begin{eqnarray}
\label{varact}
\delta \mathcal{I}&=& -2\pi (t_2-t_1)\left\{ \tilde{N}\left[g_1-\frac{2g_2+g_3g_4^2}{r}\,f' +\frac{g_3}{4r^3}f(f'-r f'')
\right] \delta f
\right. \nonumber \\[2mm]  & & \left.
-\frac{g_3}{4r^3}\left[\tilde{N} f(f'-r f'')\right]' \delta f
+\frac{g_3}{4r^2}\,\tilde{N}f\left(f'-r f''\right)\delta f'
\right. \nonumber \\[2mm]  & & \left.
-g_0\tilde{N}\delta\left(\frac{{N^r}^{\,2}}{f\tilde{N}^2}\right)
+g_0 N^\phi \delta \left(\frac{r^3{N^\phi}'}{\tilde{N}}\right)\right\}+\delta\mathcal{B}\,.
\end{eqnarray}
Clearly,  this quantity must vanish under extremizing  of the action  with appropriate boundary conditions. This implies adjusting  the boundary term  $ \mathcal{B} $  in such a way  that to cancel all the preceding terms in (\ref{varact}). With this in mind and  demanding that the fields at infinity  are determined by the solution in (\ref{stheg}), we find that the boundary term is given by
\begin{equation}
\label{B}
\mathcal{B}=(t_2-t_1)\left\{-\tilde{N}_\infty\left(2\pi m \left[g_1+ 2\left(2 g_2+g_3 g_4^2\right)\Lambda_W\right]
+\frac{2\pi g_0}{\tilde{N}_\infty^2}\,\xi\right)
+N^\phi_\infty\left(-\frac{4\pi g_0}{\tilde{N}_\infty}
\,\gamma \right)\right\}+\mathcal{B}_0,
\end{equation}
where $ B_0 $ is an arbitrary constant and we have  renamed the constants of integration $ \tilde{N}_0 $ and  $ \sigma $ in  solution (\ref{stheg})   as asymptotic displacements $ \tilde{N}(\infty)$ and $ N^\phi(\infty)  $, respectively. From this expression, we see that  the mass $ \mathcal{M} $ and the angular momentum  $ \mathcal{J} $ appear as conjugates to these asymptotic displacements, as it must be in the Hamiltonian approach under consideration. Therefore, we have
\begin{eqnarray}
\mathcal{M} &=& 2\pi m \left[g_1+ 2\left(2 g_2+g_3 g_4^2\right)\Lambda_W\right] + \frac{2\pi g_0}{\tilde{N}_\infty^2}\,\xi + C
\,,~~~~~\mathcal{J}=-\frac{4\pi g_0}{\tilde{N}_\infty}\,\gamma\,,
\label{MJ1}
\end{eqnarray}
where the appearance of an arbitrary constant $ C $ in the expression for the mass is induced by the constant $ B_0 $ present in (\ref{B}). We can now set  $ \tilde{N}(\infty)=1 $ and $ N^\phi(\infty) =0  $, without loss of generality.

Substituting  into these expressions  the quantities given in (\ref{g0123}),  with the emergent relations (\ref{emergent}) in mind,  and choosing  the constant $ C $ so as to obtain  zero mass for the disappearing event horizon (see Eqs.(\ref{btzhor1}) and (\ref{stbtzhor1})), we find that the mass and the angular momentum, per unit length  of the rotating black string   (\ref{stheg}),  are given by
\begin{eqnarray}
\mathcal{M} &=& \frac{m+\xi}{4 G} \,\,,~~~~~~~\mathcal{J}= - \frac{\gamma}{4 G}\,\,.
\label{MJ2}
\end{eqnarray}
As it was mentioned above, from these expressions it follows that the quantities  $ m +\xi $  and $ \gamma $  can be thought of as the  mass  and the rotation parameters, respectively.

\subsection{Lemos type solutions}

We turn now to the solutions for which the function $ g(r) $ in  metric (\ref{metansatz})  is given as $ g= \alpha^2 r^2 $.  Unfortunately, for the nonvanishing shift vector we were unable to solve the field equations even in the relativistic limit  $ \lambda =1 $. Therefore, we  restrict ourselves to the static case with zero shifts, but with any value of $ \lambda > 1/3 $. With these in mind, we substitute the metric ansatz  (\ref{metansatz}) into the equations of motion. As a consequence,  we find that  the momentum constraint  ({\ref{mom}}) is trivially fulfilled, while the Hamiltonian constraint ({\ref{ham}})  gives the equation
\begin{eqnarray}
&&(2\lambda-1)\,\frac{f^2}{r^2} -2\lambda\,\frac{ff'}{r}
+\frac{\lambda-1}{2}\,f'^{\,2}-2\Lambda_W(r f)'-3\Lambda_W^2 r^2 = 0\,.
\label{hamg1}
\end{eqnarray}
It is also straightforward to show that for the components of the tensor $ E_{ij} $ in equation (\ref{dyneq}), the relation $ E_{zz}= \alpha^2  E_{\phi\phi} $ holds. Therefore, we have only two independent components of equations (\ref{dyneq}). These are given by
\begin{eqnarray}
&&\left(\ln \tilde{N}\right)'\left[(\lambda-1) f'-2\lambda\,\frac{f}{r}-2\Lambda_W  r \right]+(\lambda-1)\left(f''-2\,\frac{f}{r^2}\right)
=0\,,~~~~(E_{rr}=0)\,,
\label{deqg1}
\end{eqnarray}
\begin{eqnarray}
&& \tilde{N}^{-1}\tilde{N}''\left[(\lambda-1)f'-2\lambda\, \frac{f}{r}-2\Lambda_{W} r \right]-\frac{\left(\ln \tilde{N}\right)'}{2r^2 f}\left\{3 r^2 f'\left[(1-\lambda)f' + 2\Lambda_W\,r
\right]
\right. \nonumber \\[2mm]  & & \left.
+2  f \left[5\lambda\, r f'-2 f + 2(1-\lambda) \,r^2 f'' +2\Lambda_W r^2\right]\right\}
-\frac{1}{r^3 f}\left\{(3-2\lambda)f^2 + \Lambda_W r^4
\left(f'' + 3\Lambda_W\right)
\right. \nonumber \\[2mm]  & & \left.
+r^2 f'\left[\lambda f'+2(1-\lambda)\,r f'' +2\Lambda_W r \right]
+r f \left[2(\lambda-2)f'+\lambda r f'' +(1-\lambda)r^2 f'''\right]\right\}= 0\,, ~~(E_{\phi\phi}=0)\,. \nonumber\\
\label{deqg2}
\end{eqnarray}
We note that in obtaining equation (\ref{deqg1}) we have used equation (\ref{hamg1}).  Next, introducing a new radial function $ {\cal F}(r) $  through the relation
\begin{equation}
\label{fF}
f=-\Lambda_W r^2- {\cal F}(r)\,,
\end{equation}
we put  equations  (\ref{hamg1}) and  (\ref{deqg1}) in the form
\begin{eqnarray}
\label{redeq1}
&&\frac{\lambda-1}{2}\,  {\cal F'}^2 -2\lambda\frac{{\cal F} {\cal F'}}{r}+(2\lambda-1)\,\frac{{\cal F}^2}{r^2}=0\,,\\[2mm]
&&(\ln \tilde{N})'\left[(\lambda-1) {\cal F'}-2\lambda\,\frac{ {\cal F}}{r}\right]+(\lambda-1)\left({\cal F}''-2\frac{{\cal F}}{r^2}\right) = 0\,.
\label{redeq2}
\end{eqnarray}
These equations  as well as equation (\ref{deqg2}) admit the trivial solution $ {\cal F} =0 $, leaving unconstrained the function $ \tilde{N} $. Furthermore, we have two other solutions  given by
\begin{equation}
f = -\Lambda_W r^2-  M  r^{p}\,,~~~~\tilde{N}=\tilde{N}_0 r^{1-2{p}}\,,
\label{flem}
\end{equation}
where $ M $ and $ \tilde{N}_0 $ are constants of integration and
\begin{equation}
\label{p}
p=\frac{2\lambda\pm\sqrt{6\lambda-2}}{\lambda-1}\,.
\end{equation}
The associated spacetime metric is given by
\begin{equation}
ds^2=- \tilde{N}^2f dt^2+f^{-1} dr^2 + r^2 d\phi^2+  \alpha^2 r^2 dz^2\,.
\label{strlem}
\end{equation}
We are interested in the solution that has a clear physical  meaning  in the relativistic limit $ \lambda=1 $. This corresponds to the lower sign in (\ref{p}) with  $ \lambda\in(1/3,\infty)\, $ or  $ p\in(-1,2) $. Evaluating the scalar curvature for this solution, we find the expression
\begin{equation}
R= 2 \Lambda_W \left(11-12 p + 4p^2\right) + 3 M r^{p-2} \left(2-2 p + p^2\right)\,,
\label{sccurlem}
\end{equation}
which clearly shows that at $ r=0 $ there exists a curvature singularity. It is also easy to see that for this solution  the radius of the event horizon is given by
\begin{equation}
r_+= \left[\frac{M}{-\Lambda_W}\right]^{\frac{1}{2-p}}\,,
\label{lemhor}
\end{equation}
where the parameter $ M $ is supposed to be positive and it is related to the mass per unit length of the string, see Eq.(\ref{massentr2}). Meanwhile, for the Hawking temperature evaluated by means of formulae (\ref{ht1}) we find
 \begin{equation}
T=- \frac{\tilde{N}_0 (2-p)}{4\pi}\,\,r_+^{2(1-p)} \,\Lambda_W\,.
\label{lemtemp}
\end{equation}
On the other hand, for  $ \lambda=1 $ (or $p$=1/2) we  have  solution (\ref{strlem}) with
\begin{equation}
\label{flemf}
f=-\Lambda_W r^2- M \sqrt{r}\,,~~~~~~\tilde{N} = \tilde{N}_0\,,
\end{equation}
where  $ \tilde{N}_0 $ can be set equal to one. We note that these results are in agreement with those obtained  in  \cite{cco1}  for topological black holes.

\section{Thermodynamics}

One of the most striking properties of black holes in general relativity is that they obey the laws of thermodynamics and have an entropy which is always given by one quarter of the  horizon area. However, the simple area law breaks down  for black holes in higher derivative gravity theories  \cite{lw}. Recently, this question was also raised  in the context of HL gravity \cite{cco1, cco2}. In particular, it was shown that the  entropy of spherically symmetric black holes (as well as  the topological ones) in HL gravity involves a logarithmic term, in addition to the leading ``one quarter of area" term. The logarithmic term disappears  only  for black holes for which  the scalar curvature of two-dimensional Einstein space vanishes. This fact  motivates us to study the area law for the black string configurations as well. In this section, we calculate the thermodynamical  quantities and study the area law for both  the static BTZ  and Lemos types  black string  solutions, given in (\ref{btzstat}) and (\ref{strlem}), respectively.

The thermodynamical properties of the black string configurations in HL gravity can be discussed in a similar way  to those of black holes in general relativity, using the Euclidean path integral approach \cite{gh}. Within this approach, the free energy  $ F $ of a thermodynamical ensemble  divided by the temperature $ T $  is identified with  the Euclidean action evaluated on the Euclidean continuation of the black hole  solutions. Thus, keeping in mind that  in our case  all the related quantities are taken per unit length of the black string, we   have
\begin{equation}
\label{free}
\mathcal{I}_E=\frac{F}{T}= \frac{\mathcal{M}}{T} - \mathcal{S}\,,
\end{equation}
where $ \mathcal{S} $ denotes the  entropy of  the system and the Euclidean action is related to that given in (\ref{actHAM})  as $ \mathcal{I}_E=-i\mathcal{I} $. We recall that  we are interested in the static case with  $ N_r=0 $. Therefore,  passing to the imaginary time  $ \tau = i t $   and using equation  (\ref{actden}), we obtain that
\begin{equation}
\mathcal{I}_E= 2\pi \beta \int^\infty_{r_+} dr~\tilde{N}\sqrt{f}\mathcal{H}+ \mathcal{B}_E\,,
\label{acteuc}
\end{equation}
where $ \beta=\tau_2-\tau_1 $ is the period of the Euclidean ``time" that in turn  determines the  Hawking temperature
\begin{equation}
T= \beta^{-1} =\left.\frac{\tilde{N} f'}{4\pi}\right\vert_{r_+},
\label{temp1}
\end{equation}
implying the absence of singularities at the black string horizon. An equivalent  definition is given in (\ref{ht1}) as well. Since for the static black string  solutions under consideration we have $ \mathcal{H}=0 $,
the boundary term $\mathcal{B}_E $ plays a crucial role in the variation of the action. This term must be adjusted so as to provide a true extremum of the action on these solutions. When performing the variation, as in the case of the  mass and angular momentum calculations described in the previous section, one must allow changes in the corresponding field variables which contribute to the boundary term, while keeping fixed their conjugates. In our case, the conjugate is the temperature which we keep fixed under the variation.

We first begin with the static BTZ type metric  (\ref{btzstat}) for which the Euclidean  action (\ref{acteuc}), as follows from (\ref{actden}),  can be written in the form
\begin{equation}
\mathcal{I}_E= 2\pi \beta \int^\infty_{r_+} dr~\tilde{N}\left\{g_1 \left(3\Lambda_W r+f'\right)-\frac{2 g_2+g_3 g_4^2}{2 r}\,f'^{\,2}-\frac{g_3}{8 r^3} f (f'-r f'')^2\right\}
+ \mathcal{B}_E\,.
\label{acteucbtz}
\end{equation}
The extremum of this action  on   metric (\ref{btzstat}) enables us  to fix the variation of the boundary term as
\begin{eqnarray}
\delta \mathcal{B}_E &=&  - 2\pi \beta \left\{ \tilde{N}\left[g_1-\frac{2g_2+g_3g_4^2}{r}\,f' +\frac{g_3}{4r^3}f(f'-r f'')
\right] \delta f
\right. \nonumber \\[2mm]  & & \left.
-\frac{g_3}{4r^3}\left[\tilde{N} f(f'-r f'')\right]' \delta f
+\frac{g_3}{4r^2}\,\tilde{N}f\left(f'-r f''\right)\delta f'\right\}^\infty_{r_+}\,.
\end{eqnarray}
With the BTZ type solution in (\ref{btzstat}), evaluating  the boundary term  at infinity, we find that
\begin{equation}
\mathcal{B}_E(\infty)= 2\pi \beta m \left[g_1+2(2g_2+g_3g_4^2)\Lambda_W \right] +\mathcal{B}_1\,.
\label{Bbtzinf}
\end{equation}
Meanwhile, for the boundary term at the horizon, similar calculations  yield
\begin{equation}
\mathcal{B}_E(r_+)=8 \pi^2 r_{+} \left[g_1+2(2g_2+g_3g_4^2)\Lambda_W \right] +\mathcal{B}_2\,.
\label{Bbtzhor}
\end{equation}
Here  $\mathcal{B}_1 $  and  $\mathcal{B}_2 $  are constants of integration and in obtaining (\ref{Bbtzhor}) we have used equation (\ref{temp1}) along with the fact that
\begin{equation}
\left(\delta f\right)_{r_+}=-\left(f\right)'|_{r_+}\,\delta r_+\,.
\label{dfhor}
\end{equation}
Since the on-shell value of the Euclidean action is determined by the  boundary term alone, $ \mathcal{I}_E = \mathcal{B}_E(\infty)-\mathcal{B}_E(r_+)\, $, then comparing this result with equations (\ref{free})  and  (\ref{temp1}), it is not difficult to see that the mass  and the entropy (per unit length) of the static BTZ type black string are given by
\begin{eqnarray}
\mathcal{M} & = & 2\pi  m \left[g_1+2(2 g_2+g_3 g_4^2)\Lambda_W \right],~~~~~
\mathcal{S} = 8\pi^2 r_+ \left[g_1+2(2 g_2+g_3 g_4^2)\Lambda_W\right],
\label{massentr1}
\end{eqnarray}
where we have omitted an arbitrary constant of integration, requiring that  $\mathcal{I}_E =0 $  for $ r_+\rightarrow 0 $. We note that the expression for the mass is precisely the same as that given in  (\ref{MJ1}) for $\xi=0 $ and
$ C=0 $. Remarkably, these expressions clearly delineate the contribution from higher-derivative terms in the action through the combination of constants $ g_2, \,g_3 $ and   $ g_4 $. Moreover, for $\lambda=1 $ using the value of these constants given in (\ref{g0123}), it is easy to show that $ g_1= 2(2 g_2+g_3 g_4^2)$. That is, the higher-derivative contributions to the mass and entropy result in the doubling of the ordinary Einstein-Hilbert contribution. As a consequence, we find that
\begin{eqnarray}
\mathcal{S}& = & \frac{\mathcal{A}}{2G}\,\,,
\label{entropy1}
\end{eqnarray}
where $ \mathcal{A} = 2\pi r_+ $ is the area of the horizon per unit length and we have also used the emergent relations in (\ref{emergent}). Thus, the entropy of the static BTZ type  black string in HL gravity is one half of its  horizon area. It is easy to check  that the entropy, with the mass and with the temperature  given in (\ref{massentr1}) and  (\ref{temp1}), respectively, satisfies  the first law of thermodynamics
\begin{eqnarray}
d\mathcal{M} &= & T d\mathcal{S}\,.
\label{flaw}
\end{eqnarray}

Next, we turn to the Lemos type black string  solution (\ref{strlem}) with the metric functions given in (\ref{flem}). For this solution the Euclidean action is given by
\begin{eqnarray}
\mathcal{I}_E &=& 2\pi \beta \alpha \int^\infty_{r_+} dr~\tilde{N}\left\{ g_1 \left[2(r f)' + 3\Lambda_Wr^2\right]-\frac{4 g_2}{r^2}\left(r f\right)'^{\,2}
-\frac{g_3 g_4^2}{2 r^2} \left[4f(r f)'+3 r^2 f'^{\,2}\right]\right\}
+\mathcal{B}_E\,.\nonumber\\
\end{eqnarray}
Again, the variation of the boundary term  can be found from the extremum of this action on solution (\ref{strlem}). We have
\begin{eqnarray}
\delta \mathcal{B}_E &= & - 2\pi \beta \alpha
\left\{\tilde{N}\left[2 g_1 r-(8 g_2+3 g_3 g_4^2) f'
-\frac{2(4g_2+g_3g_4^2)}{r}\,f\right]\delta f\right\}^\infty_{r_+}\,.
\end{eqnarray}
Performing similar calculations, as in the case of BTZ type black string, and comparing  the on-shell value of the Euclidean action $ \mathcal{I}_E = \mathcal{B}_E(\infty)-\mathcal{B}_E(r_+)\, $ with that given in  (\ref{free}), we find that the mass and the entropy of the Lemos type black string are given by
\begin{eqnarray}
\label{masslem2}
\mathcal{M} &=& \pi\alpha\tilde{N}_0  M^2 \left[2\left(4g_2+g_3g_4^2\right)+ p \left(8g_2+3g_3g_4^2\right)\right],\\[3mm]
\mathcal{S}&= & 4 \pi^2  r_{+}^2 \alpha \left[2 g_1+(2-p)\left(8 g_2+3 g_3 g_4^2\right)\Lambda_W\right].
\label{entrlem2}
\end{eqnarray}
We note that the mass is determined only  by the contribution from higher-derivative terms. On the other hand, it is curious that in both expressions the second term in the square bracket disappears in the relativistic limit  $\lambda= 1 $. Indeed, using equation  (\ref{g0123}), it is easy to show that the combination $ 4 g_2+g_3 g_4^2 $ vanishes identically. That is, for $\lambda= 1 $,  the entropy must obey the usual area  law just as for black holes in general relativity. Indeed, using  relations (\ref{g0123}) and (\ref{emergent}) in  equations (\ref{masslem2}) and (\ref{entrlem2}), we find that
\begin{eqnarray}
\mathcal{M} &=& -\frac{\alpha\tilde{N}_0  M^2\sqrt{6\lambda-2}}{16 G \Lambda_W}\,\,,~~~~~~ \mathcal{S}=  \frac{\sqrt{6\lambda-2}}{2}\,\frac{\mathcal{A}}{4G}\,\,,
\label{massentr2}
\end{eqnarray}
where $ \mathcal{A} =2\pi \alpha r_{+}^2 $ is  the  horizon area per unit length. We see that, for  $\lambda= 1 $, the entropy of the Lemos type black string in HL gravity is given by one quarter of the  horizon area.
It is  straightforward to verify that  with the temperature given in (\ref{lemtemp}), the quantities in (\ref{massentr2}) fulfil the first law of thermodynamics, see Eq. (\ref{flaw}).

\section{Conclusion}

In this paper, we have shown that HL gravity admits a class of cylindrically  symmetric solutions  which can be interpreted as  counterparts of black strings in general relativity. Using the general stationary and  cylindrically symmetric ansatz for the spacetime metric and focusing on the cases when  the Cotton tensor in the HL action vanishes, we have distinguished two examples of the cylindrically  symmetric spacetimes.  In the first example,  the metric  ansatz matches, for  the vanishing radial shift in the ADM-type decomposition, the form of the stationary BTZ black string metric in general relativity.  On this ground,  one can think of  the resulting solutions as  describing  the BTZ type black strings in HL gravity. In the second example, the metric ansatz corresponds to the Lemos  type  black string  configuration which  does exist in general relativity with a negative cosmological constant.

For the relativistic value of the coupling constant, $\lambda=1 $, we have given the static  BTZ type black string solutions with both zero and nonzero radial shift. The solution with the radial shift, the hedgehog type solution,  is inherent in HL gravity alone, as  the foliation-preserving invariance of the theory  is not enough to  eliminate the shift from the metric. Moreover, unlike general relativity, HL gravity provides a natural place for the BTZ type black string configurations, due to its particular higher derivative structure. As is known, in general relativity such configurations require a specific source term  for the Einstein field equations. We have also found the stationary and cylindrically  symmetric solution with the radial shift, which corresponds to a rotating BTZ type black  string.  It is important to note this solution requires the presence of the radial hair for its very existence. In other words, the radial hair is necessary for rotation.

With the Lemos type black string, restricting ourselves to the static case with zero shifts, we have presented the exact solution for any value of $ \lambda > 1/3 $.  Further, exploring the thermodynamical properties of the black strings in the framework of the Euclidean path integral approach,
we have shown that  for $\lambda=1 $ the entropy (per unit length) of the Lemos type static black string is   one quarter of the  horizon area. Meanwhile, the corresponding  entropy of the static BTZ type black string is equal to one half of its  horizon area.

\section{ACKNOWLEDGMENT}

\c{C}. \c{S}.  thanks the Scientific and Technological Research Council of Turkey (T{\"U}B\.{I}TAK) for financial support under the Programme BIDEB-2218.

\end{document}